\documentclass{jfm}
\usepackage{graphicx, color}
\usepackage{epstopdf, epsfig}
\usepackage{mathtools}
\usepackage{amsmath}
\usepackage{amssymb}
\usepackage{bm,bbm}
\usepackage{float}
\usepackage{color}
\usepackage{xcolor}
\usepackage{enumerate}
\usepackage{comment}
\usepackage{enumitem}

\newcommand{\beq}{\begin{equation}}
\newcommand{\eeq}{\end{equation}}

\renewcommand{\d} {\mathrm{d}}   

\newcommand{\Rel}{{\rm Re}_\lambda}
\newcommand{\eps}{\varepsilon}
\renewcommand{\vec}[1]{\bm{#1}}

   \newcommand{\tr}[1] {{\rm tr}\left(  {#1} \right)}      
 \renewcommand{\tr}[1] {{\rm tr}\left\{ {#1} \right\}}

  \newcommand{\Sbaromij}
     { \overline{S}^\ell_{\omega,ij}}       

\shorttitle{Helicity fluxes in homogeneous turbulence}
\shortauthor{D. Capocci, P. Johnson, S. Oughton, L. Biferale, M. Linkmann}

\title{New Exact Betchov-like Relation for the Helicity Flux in Homogeneous Turbulence}

\author{Damiano Capocci\aff{1}
\corresp{\email{capocci@roma2.infn.it}},
  Perry L. Johnson\aff{2},
  Sean Oughton\aff{3}, \\
  Luca Biferale\aff{1}
 \and Moritz Linkmann\aff{4}\corresp{\email{moritz.linkmann@ed.ac.uk}}}

\affiliation{\aff{1}Department of Physics and INFN, University of Rome Tor Vergata, Rome, Italy
\aff{2} Department of Mechanical and Aerospace Engineering,  University of California, Irvine, USA
\aff{3} Department of Mathematics, University of Waikato, Hamilton, New Zealand
\aff{4} School of Mathematics and Maxwell Institute for Mathematical Sciences, \\ University of Edinburgh, Edinburgh, EH9 3FD, United Kingdom
}

\begin{document}

\maketitle

\begin{abstract}
	In homogeneous and isotropic turbulence, the relative contributions of different physical
	mechanisms to the energy cascade can be quantified by an exact
	decomposition of the energy flux (P. Johnson, Phys. Rev. Lett., 124,
	104501 (2020), J. Fluid Mech. 922, A3(2021)). We extend the formalism 
	to the transfer of kinetic helicity across scales, important in the presence of large-scale mirror 
	breaking mechanisms, to identify physical processes resulting in helicity transfer and quantify their
	contributions to the mean flux in the inertial range. All
	subfluxes transfer helicity from large to small scales. About 50$\%$
	of the mean flux is due to the scale-local vortex flattening and vortex twisting. 
	We derive a new exact relation
	between these effects, similar to the Betchov relation for the energy flux, 
	revealing that the mean contribution of the
	former is three times larger than that of the latter.  Multi-scale
	effects account for the remaining 50$\%$ of the mean flux, with approximate
	equipartition between multi-scale vortex flattening, twisting and entangling.

\end{abstract}


\section{Introduction}

The kinetic helicity, defined as the $L^2$-inner product of 
    velocity   $ \bm{u}$ 
and vorticity  $ \bm{\omega}$, 
has dynamical, topological, geometrical, and statistical interpretations in turbulence.  It is a dynamical and topological inviscid invariant, where the latter refers to its connection with the linking number of infinitesimal vortex lines \citep{Moffatt1969
}. 
Geometrically, it quantifies the alignment of velocity and vorticity in a volume-averaged sense. Within a statistical approach to turbulence, helicity is the correlation between velocity and vorticity. In a rotationally invariant ensemble, it is connected to the breaking of the symmetry under inversion of all axes. Inspired by its relevance to turbulence in atmospheric flows \citep{Lilly86}, dynamical and statistical effects connected with helicity have been studied in the atmospheric boundary layer \citep{Deusebio14} and in rotating turbulence \citep{Mininni10a,Mininni10b}, and more generally in 
homogeneous and isotropic turbulence \citep{Chen03a,Chen03b,Gledzer15,Kessar15,Sahoo15,Stepanov15a,Alexakis17,
sahoo.prl.2017,
Milanese2021,Yan2020}, as well as shear flows \citep{Yan2020, Yu2022} and in laboratory experiments \citep{Scheeler2017}.

The level of helicity in a turbulent flow affects turbulent statistics and
dynamics, and is thus of relevance from a fundamental theory perspective as
well as for subgrid-scale (SGS) modelling. As an alignment of velocity and
vorticity weakens the nonlinearity of the Navier--Stokes equations, high levels
of helicity have been connected with a depletion of the kinetic energy flux
across scales by an analysis of the coupling between helical Fourier modes
\citep{Kraichnan73}, and with regions of low dissipation \citep{Moffatt2014a}.
These 
effects can be quantified by upper bound theory applied
to helical forcing and direct numerical simulation --- the
energy flux of turbulence sustained by fully helical forcing is about $30 \%$
lower than in the non-helical case \citep{Linkmann2018}.

Helicity affects turbulence not only globally, that is, in terms of 
  \emph{mean} energy
fluxes, but also on a scale-by-scale level. 
As a solenoidal vector field, the velocity field $\bm{u}$ can be decomposed into positively and negatively helical 
components $\bm{u}^\pm$ \citep{Herring74,Constantin88,Waleffe92},
$    \bm{u}(\bm{x},t) =\bm{u}^+(\bm{x},t) + \bm{u}^-(\bm{x},t) $, where $\bm{u}^\pm$ are obtained
by projecting the Fourier coefficients $\hat{\bm{u}}(\bm{k},t)$ 
onto basis vectors which are eigenfunctions of the curl operator in Fourier space. That is, 
$
	\hat{\bm{u}}^\pm(\bm{k},t) = u^\pm (\bm{k},t) \bm{h}^\pm(\bm{k}) \ ,
$
where $i \vec{k} \times k\bm{h}^\pm(\bm{k}) = \pm \bm{h}^\pm(\bm{k})$ and $u^\pm (\bm{k},t) = \hat{\bm{u}}(\bm{k},t) \cdot \bm{h}^\pm(\bm{k})$. 
The energy flux can then be decomposed into different triadic
couplings between positively and negatively helical velocity-field fluctuations
\citep{Waleffe92}. Interestingly, interactions among helical Fourier modes of
like-signed helicity leads to an inverse energy transfer across scales in the
inertial range \citep{Waleffe92,Biferale2012,Biferale2013a, Sahoo15
},
while interactions of oppositely-signed helical modes transfer energy from
large to small scales \citep{Waleffe92, Alexakis17, AlexakisBiferale2018}. 
For turbulent flows of electrically conducting fluids such as liquid metals or plasmas in the fluid approximation, helicity alters the evolution of both velocity and magnetic-field fluctuations profoundly. Here, small-scale kinetic helicity facilitates  the formation of
large-scale coherent magnetic structures through the large-scale dynamo 
\citep{Steenbeck66,Brandenburg01,Brandenburg05,Tobias13,Linkmann2016,Linkmann2017}.

The cascade of kinetic helicity itself is predicted to be direct, that is, it proceeds
from large to small scales \citep{Brissaud73,Waleffe92}, and scale-local \citep{Eyink05}. 
It results, as discussed by \citet{Eyink06} in the context of a multi-scale gradient expansion, from a twisting of small-scale vortices into a local alignment with the small-scale velocity fluctuations by large-scale differential vorticity (`screw'). However, being sign-indefinite,
numerical results on helicity fluxes can be difficult to interpret as a loss of positive helicity at a given scale may be viewed as a gain of negative helicity at the same scale. 

%
%


In the context of SGS modelling, the effect helicity has on a turbulent flow is
usually taken into account though additional diffusive model terms
\citep{Yokoi1993,Li2006,Baerenzung2008,Inagaki2017}. However, a combination of
\emph{a-priori} and \emph{a-posteriori} analyses of different SGS models for
isotropic helical turbulence found the effect of the additional diffusive model
terms to be small and 
that a classical Smagorinsky model best represents the
resolved-scale dynamics \citep{Li2006}. 
Similarly, based on analytical and
numerical results, \citet{Linkmann2018} suggests an adjustment of the
Smagorinsky constant to account for high levels of helicity. 
So far, SGS analyses of helical turbulence have mainly been concerned with
energy transfers.  

Here, we focus on the helicity flux across scales in
statistically stationary homogeneous and isotropic turbulence, with large-scale forcing breaking mirror symmetry. 
For the energy flux, the 
    \cite{Betchov56} relation 
states that the mean contribution from
vortex stretching to the energy cascade is triple that due to
strain self-amplification. \citet{CarboneWilczek22} recently showed that there are
no further kinematic relations for the \emph{energy} flux in statistically stationary
homogeneous and isotropic turbulence with zero net helicity.
However, we prove here that 
a new exact  kinematic Betchov-type relation exists for the mean \emph{helicity} flux.  
Furthermore, we also present an exact decomposition of the helicity flux in analogy to
that of the kinetic energy flux derived by
\citet{Johnson2020,Johnson2021}, whereby the relative contributions of physical
mechanisms, such as vortex stretching and strain self-amplification, to the
energy cascade can be quantified in terms of the overall contribution and their
scale-locality. The aim is to
identify 
physical mechanisms that transfer kinetic helicity across
scales and to quantify their relative contributions to the mean helicity flux and its fluctuations, 
which may be useful for the construction of SGS models 
when resolving the helicity cascade is of interest.

\vspace{-1em}
\section{Exact decomposition of the kinetic helicity flux} \label{sec:decomp_helflux}

To derive the aforementioned exact decomposition of the helicity flux and relations between the resulting subfluxes, 
we begin with the three-dimensional (3D) incompressible Navier--Stokes equations, here written in component form 
\begin{align}
\label{eq:momentum}
    \partial_t u_i + \partial_j\left (u_i u_j \right) 
    & = - \partial_j p \delta_{ij} + 2 \nu \partial_j S_{ij} + f_i \ , \\
    \label{eq:incomp}
    \partial_j  u_j  & = 0 \ , 
\end{align} 
where $\bm{u} = (u_1, u_2, u_3)$ is the velocity field, 
$p$ the pressure divided by the constant density, 
$\nu$ the kinematic viscosity, 
$S_{ij}$ the rate-of-strain tensor, 
and $\bm{f} = (f_1, f_2, f_3)$ an external solenoidal force that may be present. 
To define the helicity flux across scales, we introduce a filtering operation to separate large- and small-scale dynamics
      \citep[e.g.,][]{Germano92}.
Specifically, for a generic function $\phi$, the filtered version 
at scale $\ell$ 
is
$
\overline{\phi}^\ell = G^\ell * \phi \ ,
$
where $G^\ell$ is a filter kernel with filter width $\ell$ and the asterisk denotes the convolution operation. 
Applying the filter to the 
Navier--Stokes equations \eqref{eq:momentum}--\eqref{eq:incomp} results in
\begin{equation}
\partial_t \overline{u}^\ell_i + \partial_j\left( \overline{u}^\ell_i \overline{u}^\ell_j + \overline{p}^\ell \delta_{ij} - 2 \nu \overline{S}^\ell_{ij} + \tau_{ij}^\ell \right) = \overline{f}_i^\ell \ ,
\end{equation}
where $\tau_{ij}^\ell = \tau^\ell(u_i, u_j) = \overline{u_i u_j}^\ell - \overline{u}^\ell_i \overline{u}^\ell_j$ is the SGS stress tensor. 
Here, we follow the notation of \citet{Germano92} in defining the generalised second moment for any two fields as $\tau^\ell(a,b) = \overline{ab}^\ell - \overline{a}^\ell \overline{b}^\ell$.
We also require the filtered vorticity equation
\beq
\label{eq:NS_vorticity}
\partial_t \overline{\omega}^\ell_i
+ \partial_j\left( \overline{\omega}^\ell_i \overline{u}^\ell_j - \overline{u}^\ell_i \overline{\omega}^\ell_j - \nu \partial_j \overline{\omega}^\ell_i \right) 
- \overline{g}_i^\ell 
= - \partial_j\left(\epsilon_{imn} \partial_m \tau^\ell_{nj}  \right)  \ ,
\eeq
where $\bm{g} = \nabla \times \bm{f}$. 
The large-scale helicity density, $H^\ell = \overline{u}^\ell_i \overline{\omega}^\ell_i$, then evolves according to
\begin{align}
\label{eq:evolH1}
\partial_t H^\ell & + \partial_j
\left[ H^\ell \overline{u}^\ell_j + (\overline{p}^\ell - \tfrac{1}{2} \overline{u}^\ell_i \overline{u}^\ell_i) \overline{\omega}^\ell_j  - \nu \partial_j H^\ell \right] 
+ 2\nu ( \partial_j \overline{u}^\ell_i )
 ( \partial_j \overline{\omega}^\ell_i )
- \overline{\omega}_i^\ell\overline{f}_i^\ell
- \overline{u}_i^\ell\overline{g}_i^\ell \nonumber \\
 = &
- \! \partial_j \! \left[ 2 \overline{\omega}^\ell_i \tau^\ell_{ij} + \epsilon_{ijk} \overline{u}^\ell_i \partial_m \tau^\ell_{km} \right]
+ 2 \tau^\ell_{ij} \partial_j \overline{\omega}^\ell_i 
\end{align}
%
%
The last term in this equation is the helicity flux
\begin{equation}
	\label{eq:helflux1}
\Pi^{H, \ell} = -  2\tau^\ell_{ij} \partial_j \overline{\omega}^\ell_i \ ,
\end{equation}
and is the central focus herein.
It has an alternative form
  \citep{Yan2020},
\begin{equation}
	\label{eq:helflux2}
\tilde{\Pi}^{H, \ell} = - \tau^\ell_{ij} \partial_j \overline{\omega}^\ell_i - \left[ \tau^\ell(\omega_i, u_j) - \tau^\ell(u_i, \omega_j) \right] \partial_j \overline{u}^\ell_i \ , 
\end{equation}
and it can be shown that the RHSs of \eqref{eq:helflux1} and \eqref{eq:helflux2}
differ by an expression that can be written as a divergence and therefore vanishes after averaging spatially, at least for statistically homogeneous turbulence \citep{Yan2020}. 
This implies 
$\langle \Pi^{H, \ell} \rangle = \langle \tilde \Pi^{H, \ell} \rangle $.
\citet{Eyink06} links the first term in \eqref{eq:helflux2} --- which is proportional to $\Pi^{H, \ell}$ --- to vortex twisting and \citet{Yan2020} attribute the second term to vortex stretching. 
In what follows we discuss an exact decomposition of $\Pi^{H, \ell}$, and show that both effects can be identified therein. We also use 
    $ \Pi^{H, \ell} $
for our numerical evaluations \citep[cf.][]{Chen03a,Eyink06}.

    \subsection{Gaussian filter relations for the helicity flux}

So far all expressions are exact and filter-independent. To
derive exact decompositions of the helicity flux in both representations, we
now focus on Gaussian filters. For that case, \citet{Johnson2020,Johnson2021}
showed that the subgrid-scale stresses can be obtained as the solution of a
forced diffusion equation with $\ell^2$ being the time-like variable, resulting in 
\begin{equation}
\tau_{ij}^\ell = \tau^\ell(u_i, u_j) = \ell^2 \overline{A}_{ik}^\ell \overline{A}_{jk}^\ell + \int_{0}^{\ell^2} \d\theta~ 
	\tau^\phi\left( \overline{A}_{ik}^{\sqrt{\theta}}, \overline{A}_{kj}^{\sqrt{\theta}} \right)
  ,
\label{eq:tau_gradients}
\end{equation}
where $\phi(\theta) = \sqrt{\ell^2 - \theta}$, and $A_{ij} = \partial_j u_i$ are the velocity-field gradients.
Since the SGS stress tensor $\tau^\ell_{ij}$ is symmetric, for the first form of the helicity flux we obtain in analogy to the energy flux
\begin{equation}
   \Pi^{H, \ell} 
        = 
    - 2 \tau^\ell_{ij}  \Sbaromij ,
\end{equation}
where 
  $ S_\omega $ 
is the symmetric component of the vorticity gradient tensor, 
with components
 $ S_{\omega, ij} = (\partial_j \omega_i + \partial_i \omega_j)/2 $.
Employing 
  \eqref{eq:tau_gradients} 
this yields
\begin{equation}
\Pi^{H,\ell} = - 2 \ell^2 \Sbaromij \overline{A}_{ik}^\ell \overline{A}_{jk}^\ell 
- 2 \int_{0}^{\ell^2} \d\theta~ \Sbaromij 
	\tau^\phi\left( \overline{A}_{ik}^{\sqrt{\theta}}, \overline{A}_{kj}^{\sqrt{\theta}} \right)
 .
\end{equation}
The first term involves a product of gradient tensors filtered at the same scale, $\ell$; hence we refer to it as being \emph{single-scale}, and denote it
    $ \Pi_s^{H,\ell} $. In mean, it coincides with the nonlinear LES model for the SGS-stresses \citep{Eyink06}. 
In contrast, the second term encodes the  correlation between 
 resolved-scale vorticity-field gradients and 
(summed) velocity-field gradients at each scale smaller than $\ell$, 
so that we refer to it as \emph{multi-scale}.

Splitting the velocity gradient tensors into symmetric and anti-symmetric parts, that is, into the rate-of-strain tensor $S = (A + A^t)/2$ and vorticity tensor $\Omega = (A-A^t)/2$, 
where $A^t$ is the transpose of $A$, the helicity flux can be decomposed into six subfluxes
\begin{equation}
\Pi^{H,\ell} = \Pi_{s, SS}^\ell + \Pi_{s,\Omega\Omega}^\ell + \Pi_{s,S\Omega}^{\ell} + \Pi_{m,SS}^\ell + \Pi_{m,\Omega\Omega}^\ell + \Pi_{m,S\Omega}^{\ell}
 ,
\label{eq:helicity-flux-decomposition}
\end{equation}
where the single-scale terms are
\begin{align}
\label{eq:pi_l_S}
\Pi_{s,SS}^{H,\ell} & 
  = - 2 \ell^2 \Sbaromij \overline{S}_{ik}^\ell \overline{S}_{jk}^\ell 
  = - 2 \ell^2 \tr{(\overline{S}^\ell_\omega)^t \overline{S}^\ell (\overline{S}^\ell)^t}\ ,
\\
\label{eq:sSAA}
\Pi_{s,\Omega\Omega}^{H,\ell} & 
  = - 2 \ell^2 \Sbaromij \overline{\Omega}_{ik}^\ell \overline{\Omega}_{jk}^\ell 
  = -2 \ell^2 \tr{(\overline{S}^\ell_\omega)^t \overline{\Omega}^\ell (\overline{\Omega}^\ell)^t}\ ,
\\
\label{eq:pi_l_c}
\Pi_{s, S\Omega}^{H,\ell} & 
  = - 2 \ell^2 \Sbaromij \left ( \overline{S}_{ik}^\ell \overline{\Omega}_{jk}^\ell  -  \overline{\Omega}_{ik}^\ell \overline{S}_{jk}^\ell \right ) 
  = -4 \ell^2 \tr{(\overline{S}^\ell_\omega)^t \overline{S}^\ell (\overline{\Omega}^\ell)^t}\ ,
\end{align}
and $\tr{\cdot}$ denotes the trace. 
Similarly, the multi-scale terms are
\begin{align}
    \label{eq:mSSS}
	\Pi_{m,SS}^{H,\ell} & = - 2 \int_{0}^{\ell^2} \d\theta~ \Sbaromij \tau^\phi\left( \overline{S}_{ik}^{\sqrt{\theta}}, \overline{S}_{kj}^{\sqrt{\theta}} \right)
 ,
\\
 \label{eq:mSAA}
	\Pi_{m,\Omega\Omega}^{H,\ell} & 
   = 
    \;\;\;  2 \int_{0}^{\ell^2} \d\theta~ \Sbaromij \tau^\phi\left( \overline{\Omega}_{ik}^{\sqrt{\theta}}, \overline{\Omega}_{kj}^{\sqrt{\theta}} \right)
 ,
\\
 \label{eq:mSSA}
	\Pi_{m,S\Omega}^{H,\ell} 
	& =  - 2 \int_{0}^{\ell^2} \d\theta~ \Sbaromij \left[ \tau^\phi\left( \overline{S}_{ik}^{\sqrt{\theta}}, \overline{\Omega}_{jk}^{\sqrt{\theta}} \right) + \tau^\phi\left( \overline{\Omega}_{ik}^{\sqrt{\theta}}, \overline{S}_{jk}^{\sqrt{\theta}} \right) \right] 
	\nonumber \\
	& = - 4 \int_{0}^{\ell^2} \d\theta~ \Sbaromij \tau^\phi\left( \overline{S}_{ik}^{\sqrt{\theta}}, \overline{\Omega}_{jk}^{\sqrt{\theta}} \right)
 .
\end{align} 

We recall that 
   $ \langle \Pi^{H,\ell}_{s,\Omega\Omega}\rangle $,
the spatial average of the contribution to the helicity flux due to coupling of resolved-scale vorticity strain with resolved-scale vorticity,  vanishes
\beq
\label{eq:SAA_vanish}
\langle \Pi^{H,\ell}_{s,\Omega\Omega}\rangle =  - \frac{\ell^2}{4} \left \langle  \left( \partial_j \overline{\omega}^\ell_i + \partial_i \overline{\omega}^\ell_j\right)\overline{\omega}^\ell_i \overline{\omega}^\ell_j \right \rangle = 
- \frac{\ell^2}{4} \left \langle  \partial_j (\overline{\omega}^\ell_i \overline{\omega}^\ell_i \overline{\omega}^\ell_j) \right \rangle = 0 \ ,  
\eeq
due to periodic boundary conditions 
and the divergence-free nature of the vorticity field, as previously discussed by \citet{Eyink06} in the context of a multi-scale gradient expansion of the SGS stress tensor.

The physics encoded in these transfer terms may be understood 
in terms of three effects:
(i) ``vortex flattening'' -- compression and stretching of a vortex tube into a vortex sheet by large-scale straining motion, with the principal axes of the vorticity deformation tensor $S_\omega$ aligning with that of the strain-rate tensor at smaller scale, see 
\eqref{eq:pi_l_S} and \eqref{eq:mSSS};
(ii) ``vortex twisting'' -- a twisting of small-scale vortex tubes by large-scale differential vorticity into thinner tubes consisting of helical vortex lines, and subsequent small-scale alignment between the resulting vorticity vectors and the extensile stress generated thereby \citep{Eyink06}, see 
\eqref{eq:pi_l_c} and \eqref{eq:mSSA};
and 
(iii) ``vortex entangling'' -- twisting of entangled vortex lines, see 
\eqref{eq:sSAA} and \eqref{eq:mSAA}.
Interpreting helicity as the correlation between velocity and vorticity, a change in this correlation (or alignment) \emph{across scales} occurs by vorticity deformation through straining motions or differential vorticity. This results in decorrelation at large scales and an increase in small-scale correlation. 

\subsection{An exact Betchov-type relation for the helicity flux} \label{sec:betchov}

In homogeneous turbulence, the \cite{Betchov56} relation is an exact expression connecting the contributions associated with vortex
stretching and strain self-amplification to the mean energy flux across scales.
Here we show that there is an analogous exact expression relating two (single scale) mean helicity subfluxes: 
  $ 3\langle \Pi^{H, \ell}_{s,SS} \rangle 
      = 
     \langle  \Pi^{H, \ell}_{s,S\Omega} \rangle
  $.
These subfluxes are associated with 
vortex flattening, $\langle \Pi^{H, \ell}_{s,SS} \rangle$, and 
vortex twisting, $\langle  \Pi^{H, \ell}_{s,S\Omega} \rangle$.
Written in 
terms of the definitions 
given in 
  \eqref{eq:pi_l_S} and \eqref{eq:pi_l_c}, this expression reads
\begin{equation}
3\, \left \langle \tr{\overline{S}^\ell_\omega \overline{S}^\ell \overline{S}^\ell} \right \rangle 
= 2\, \left \langle \tr{\overline{S}^\ell_\omega \overline{\Omega}^\ell \overline{S}^\ell} \right\rangle
 .
\label{eq:betchov_trace}
\end{equation}

The main steps in a proof of this are now summarised. 
Following an argument analogous to that used in proving the   \cite{Betchov56} 
relation for the energy flux, and using tensor symmetry properties and
  \eqref{eq:SAA_vanish}, 
one obtains
  \citep{Eyink06}
\beq
\label{eq:pre_betchov}
\left \langle \tr{\overline{S}^\ell_\omega \overline{S}^\ell \overline{S}^\ell} \right \rangle = - \left \langle \tr{\overline{\Omega}^\ell_\omega (\overline{S}^\ell \overline{\Omega}^\ell + \overline{\Omega}^\ell \overline{S}^\ell) } \right \rangle = - 
2\left \langle \tr{\overline{\Omega}^\ell_\omega \overline{\Omega}^\ell \overline{S}^\ell} \right \rangle \ ,
\eeq
where $\Omega_\omega$ is the antisymmetric part of the vorticity gradient tensor.
This yields
	\begin{align}
	\label{eq:betchov_fulltensor}
		\frac{1}{2}\left \langle \tr{ \nabla \overline{\bm{\omega}}^\ell \left (\nabla \overline{\bm{u}}^\ell \right )^t 
  \left[ \left ( \nabla \overline{\bm{u}}^\ell + \nabla \overline{\bm{u}}^\ell \right )^t  \right] }\right \rangle 
	     & =  \left \langle \tr{\frac{3}{2} \,\overline{S}^\ell_\omega \overline{S}^\ell \overline{S}^\ell - \,\overline{S}^\ell_\omega \overline{\Omega}^\ell \overline{S}^\ell} \right\rangle 
	  .   
	\end{align}
Thus, showing that the lefthand side (LHS) of this expression vanishes
will prove the Betchov relation for the helicity flux, 
  \eqref{eq:betchov_trace}. 
To do so, we express the LHS of eq.~\eqref{eq:betchov_fulltensor} using the chain rule and in index notation
\begin{align}
\left \langle 
\partial_j \overline{\omega}_i^\ell \partial_j \overline{u}_k^\ell \overline{S}_{ki}^\ell 
\right \rangle
= &  \left \langle \partial_j \left [\overline{\omega}_i^\ell \partial_j \overline{u}_k^\ell \overline{S}_{ki}^\ell \right] \right \rangle 
\nonumber \\
& -
 \left \langle
\overline{\omega}_i^\ell \partial_j \partial_j \overline{u}_k^\ell \overline{S}_{ki}^\ell 
\right \rangle
-
 \left \langle
\overline{\omega}_i^\ell  \overline{S}_{kj}^\ell  \partial_j\overline{S}_{ki}^\ell 
\right \rangle
-
 \left \langle
\overline{\omega}_i^\ell  \overline{\Omega}_{kj}^\ell  \partial_j\overline{S}_{ki}^\ell 
\right \rangle \ .
\end{align}
The first term on the RHS of this expression vanishes making use of periodic boundary conditions. Using incompressibility and integration by parts it can be shown that the last term also vanishes. The two remaining terms cancel out, which is shown by similar arguments and using the properties of the Levi-Civita tensor. This completes the proof.

The mean single-scale terms also arise as the first-order contribution in a multi-scale expansion of the SGS stress tensor \citep{Eyink06}, where 
  \eqref{eq:pre_betchov} 
is used to deduce that the full vorticity gradient, not only either its symmetric or antisymmetric component, is involved in the helicity flux across scales. 
In consequence, \eqref{eq:betchov_trace} and \eqref{eq:pre_betchov} assert that the mean transfers involving the symmetric or the antisymmetric parts of the vorticity gradient 
can be related to one another, 
and thus the single-scale contribution to the  mean  helicity flux can be written as
\beq
 \left \langle \Pi^{H,\ell}_s \right \rangle
=  -8\ell^2\, \left \langle \tr{\overline{S}^\ell_\omega \overline{S}^\ell \overline{S}^\ell} \right \rangle 
= -\frac{16}{3} \ell^2\, \left \langle \tr{\overline{S}^\ell_\omega \overline{\Omega}^\ell \overline{S}^\ell} \right\rangle \ .
\eeq



\section{Numerical details and data}\label{sec:numerics}

Data has been generated by direct numerical simulation of the incompressible 3D Navier--Stokes equations 
\eqref{eq:momentum} and	\eqref{eq:incomp} on a triply periodic domain of size $L_{\rm box} = 2\pi$ in each direction, 
where the forcing $\bm{f}$ is a random Gaussian process with zero mean, fully helical $\bm{f} = \bm{f}^+$, and 
active in the wavenumber band $k \in [0.5,2.4]$. 
The spatial discretisation is implemented through the standard, fully dealiased 
pseudospectral method with $1024$ collocation points in each direction. 
Further details and mean values of key observables are summarised in table \ref{tab:simulations}.

\begin{table}
  \begin{center}
\def~{\hphantom{0}}
   \begin{tabular}{ccccccccccccc}
        \hline
        \hline
		 $N$ & $E$  & $\nu$ & $\eps$ & $\eps_H$ & $L$ & $\tau$  & $\Rel$ &  $\eta/10^{-3}$ & $k_{\rm max}$ & $k_{\rm max} \eta$ & $\Delta t / \tau$ & \# \\
        \hline
		 1024 & 7.26  & 0.001 & 3.33 & 5.02  & 1.12  & 0.50    & 327    & 4.20  & 340  & 1.43 & 0.60  & 39 \\

        \hline
      \hline
        \end{tabular}
        \caption{Simulation parameters and key observables, where
        $N$ is the number of collocation points in each coordinate,
        $E$ the (mean) total kinetic energy,
        $\nu$ the kinematic viscosity,
        $\eps$ the mean energy dissipation rate,
        $\eps_H$ the mean helicity dissipation rate, 
        $L = (3 \pi/4 E) \int_0^{k_{\rm max}} \d{k} \ E(k)/k$ the integral scale,
        $\tau = L/\sqrt{2E/3}$ the large-eddy turnover time,
        $\Rel$ the Taylor-scale Reynolds number,
        $\eta = (\nu^3/\eps)^{1/4}$ the Kolmogorov microscale,
        $k_{\rm max}$ the largest wave number after de-aliasing,
        $\Delta t$ the sampling interval which is calculated from
        the length of the averaging interval divided by the number of equispaced snapshots, and 
	$\#$ the number of snapshots.   
	The data corresponds to run 22 of ~\citet{sahoo.prl.2017}. 
	It is available for download using the SMART-Turb portal {\tt http://smart-turb.roma2.infn.it}. 
	}
  \label{tab:simulations}
  \end{center}
\end{table}

\begin{figure}
	\begin{center}
         \includegraphics[width=.4\columnwidth]{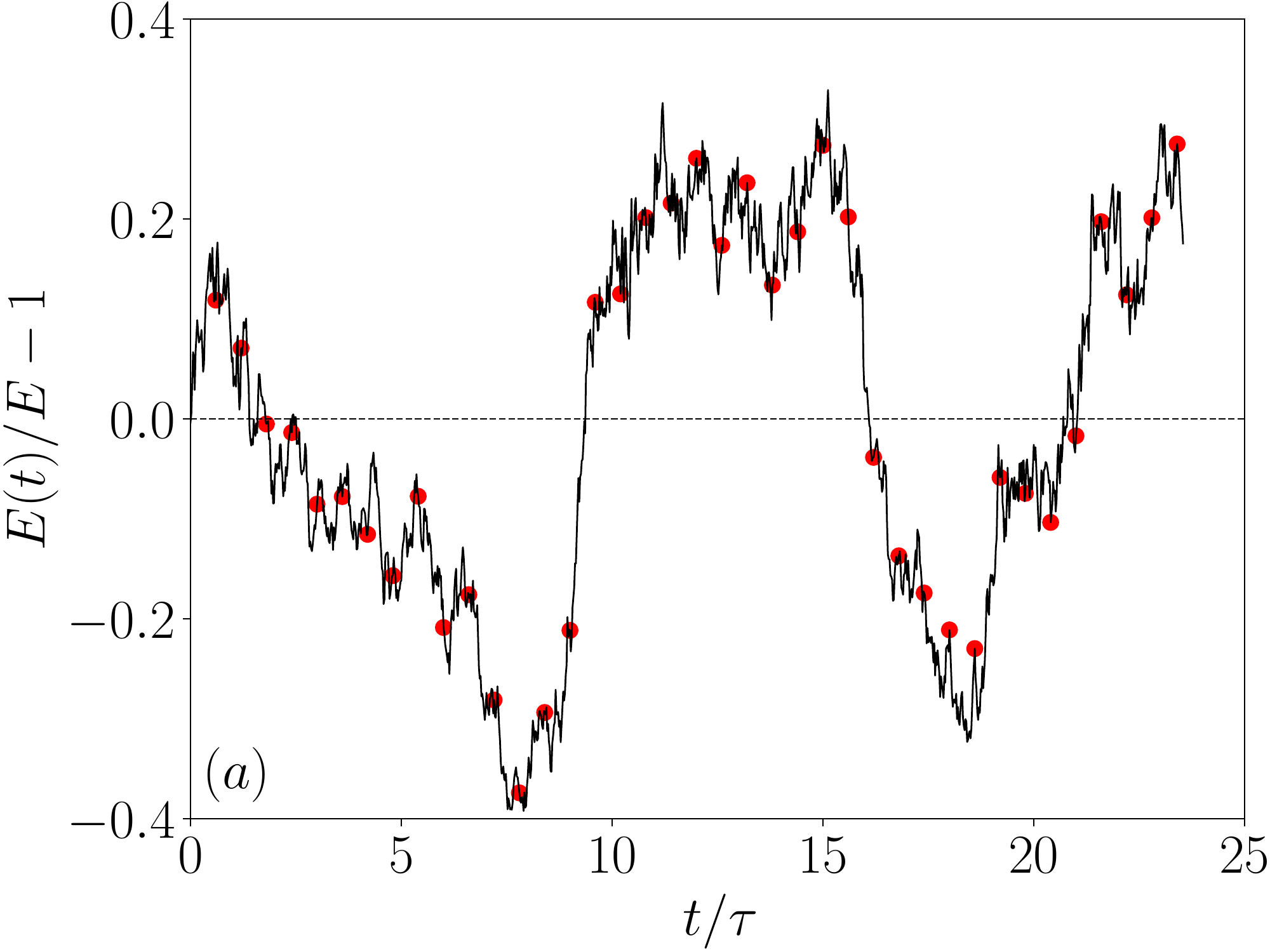} 
		\hspace{2em}
        \includegraphics[width=.4\columnwidth]{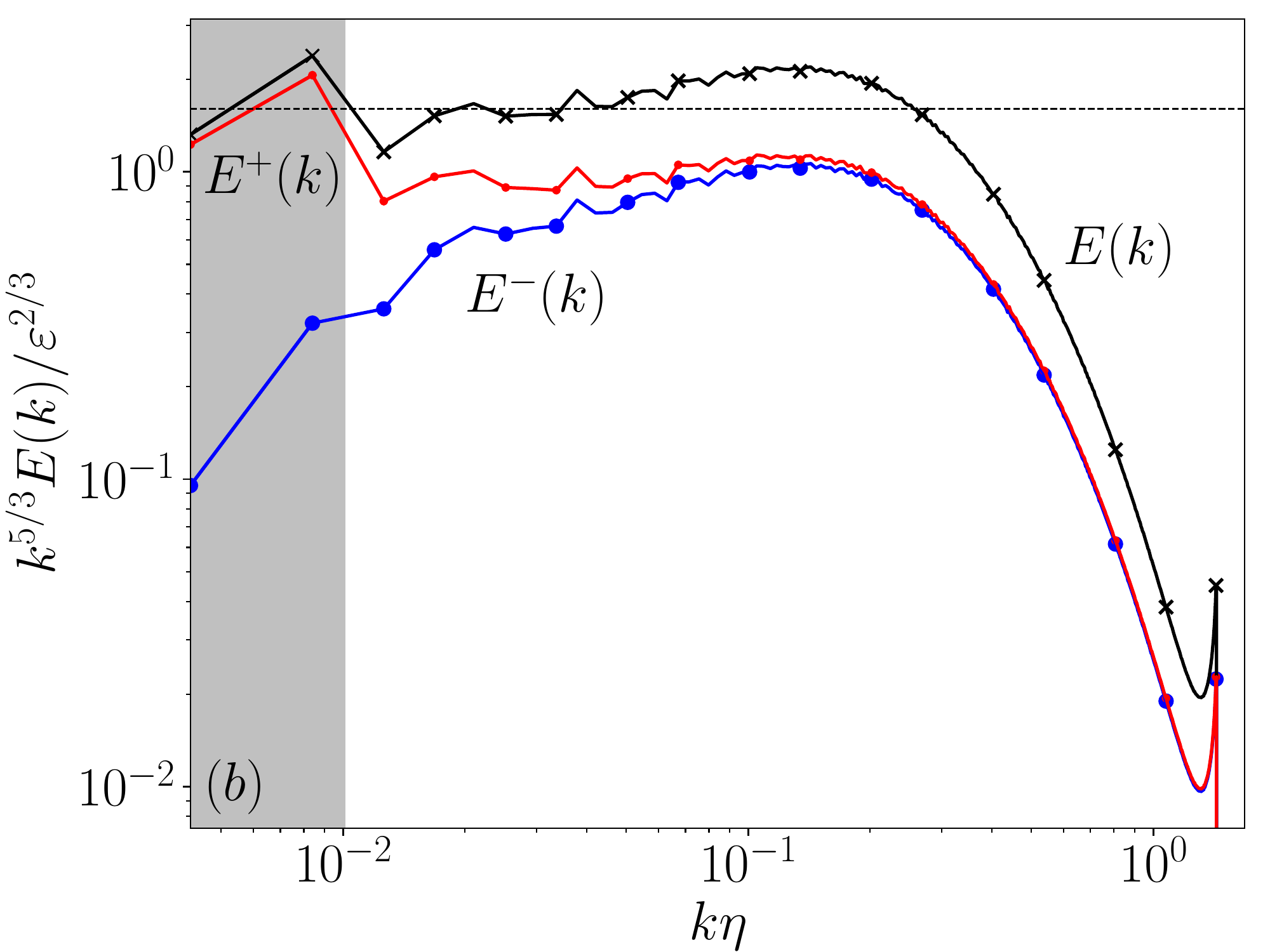}
        \end{center}
	 \caption{
	 (a) Time evolution of the total energy 
	 normalised by its mean value, $E$. 
	 Time is given in units of large-eddy turnover time $\tau$.
         The red dots correspond to the sampled velocity-field configurations.
         (b) Time-averaged energy spectra in Kolmogorov-compensated form. 
        The grey-shaded area indicates the forcing range.
	The dashed line indicates a Kolmogorov constant $C_K \approx 1.6$.
         }
	 \label{fig:timeseries}
\end{figure}

Figure \ref{fig:timeseries}(a) presents the time series of the total kinetic energy per unit volume, $E(t)$. 
Time-averaged kinetic energy spectra of positively and negatively helical fluctuations,
$E^\pm(k) = \langle \frac{1}{2}\sum_{k \leqslant |\bm{k}| < k + 1} |\hat{\bm{u}}^\pm(\bm{k})|^2 \rangle$ and 
the total energy spectrum $E(k) = E^+(k) + E^-(k)$, are shown in
in Kolmogorov-compensated form in Fig.~\ref{fig:timeseries}(b).
%
%
%
As can be seen by comparison of $E^+(k)$ and
$E^-(k)$, the large-scale velocity-field fluctuations are dominantly positively helical,
which is a consequence of the forcing. Decreasing in scale, we observe that
negatively helical fluctuations increase in amplitude, and approximate
equipartition between $E^+(k)$ and $E^-(k)$ is reached for $k \geqslant 20$.
That is, a helically forced turbulent flow, where mirror-symmetry is broken at
and close to the forcing scale, restores mirror-symmetry at smaller scales
through nonlinear interactions \citep{Chen03a,Deusebio14,Kessar15}.

%

\vspace{-1em}
\section{Numerical results for mean subfluxes and fluctuations} \label{sec:calculations}


\begin{figure}
        \includegraphics[width=.6\columnwidth]{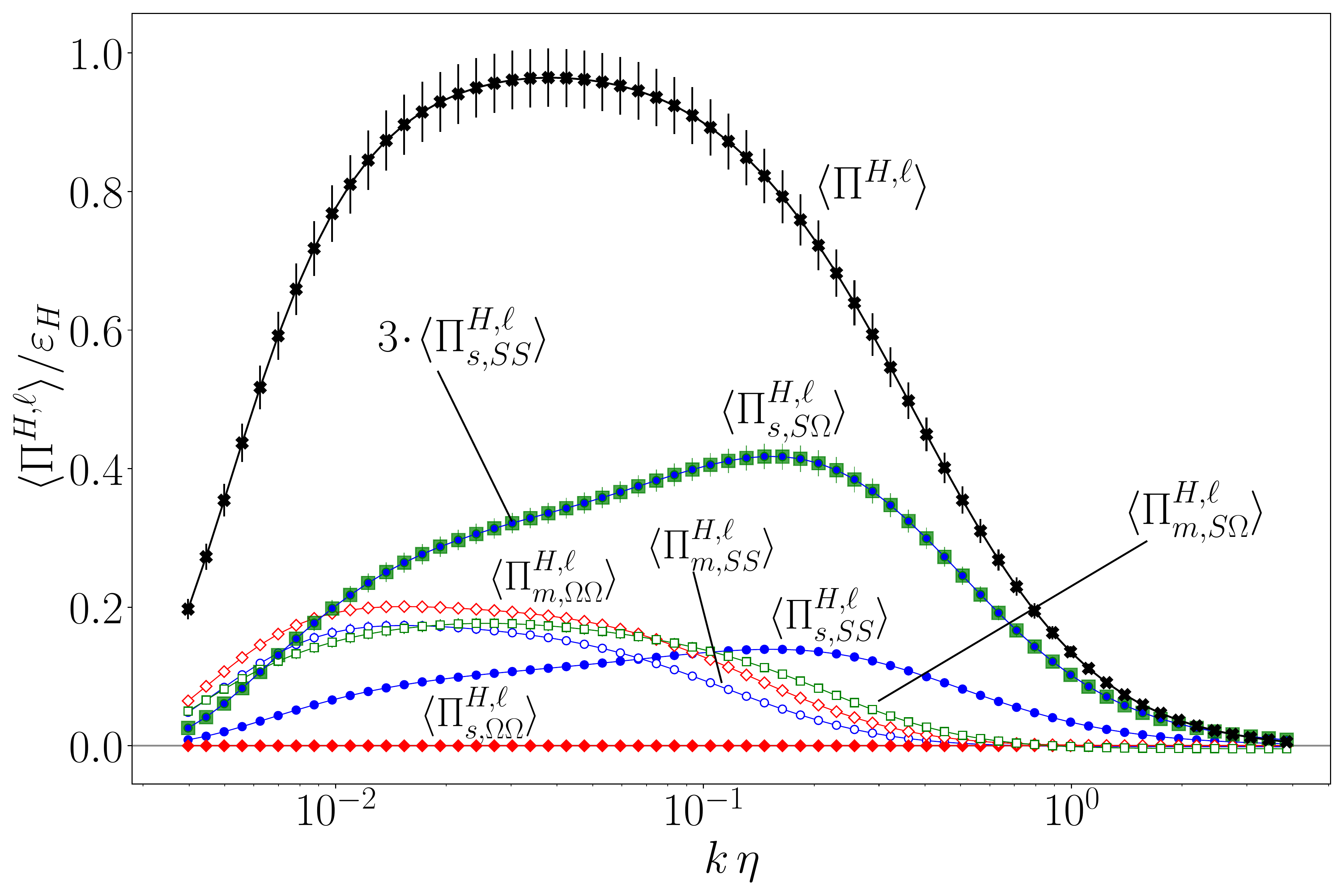}
        \centering
        \caption{
		Decomposed helicity fluxes normalised with the mean helicity dissipation rate $\eps_H$. Filled markers corresponds to single-scale contributions while empty symbols are related to multi-scale contributions. The error bars indicate one standard error. 
	 %
        The subflux $\langle\Pi_{s,S\Omega}^{H,\ell}\rangle$ has been superposed with $3\langle\Pi_{s,SS}^{H,\ell}\rangle$ in order to highlight the Betchov-type relation 
           \eqref{eq:betchov_trace}.}
	\label{fig:helicity}
\end{figure}

Figure \ref{fig:helicity} shows the total helicity flux and all subfluxes, normalised 
by the total helicity dissipation rate $\varepsilon_H$. 
As can be seen in the figure, the term $\langle \Pi^{H,\ell}_{s,\Omega\Omega}\rangle$ is identically zero, which must be the case according to 
   \eqref{eq:SAA_vanish}. 
Moreover, the helicity Betchov relation
  \eqref{eq:betchov_trace} 
derived here is satisfied as it must be -- the terms $\langle \Pi^{H,\ell}_{s,S\Omega}\rangle$ and $3\,\langle \Pi^{H,\ell}_{s,SS}\rangle$ are visually indistinguishable, with a relative error between them of order $10^{-6}$ (not shown). 
A few further observations can be made from the data.
The non-vanishing multi-scale terms, $\langle  \Pi^H_{m,S\Omega} \rangle$, $\langle  \Pi^H_{m,SS} \rangle$ and 
$\langle  \Pi^H_{m,\Omega\Omega} \rangle$ are comparable in magnitude across all scales. They are approximately scale-independent in the interval $10^{-2} \leqslant k\eta \le 10^{-1}$, with each accounting for about $15-20\%$ of the total helicity flux in this range of scales. Even though clear plateaux are not present for the two non-vanishing single-scale terms, $\langle  \Pi^H_{s,S\Omega} \rangle$ and $\langle  \Pi^H_{s,SS} \rangle$,  
one could tentatively extrapolate that at higher Re, about $30\%$ of the mean flux originates from scale-local vortex twisting and $10\%$ from vortex flattening. 
%
%
%
That is, the multi-scale contributions amount to 50\%-60\% and the scale-local contributions to 40-50\% of the
total helicity flux across scales, at least for this particular simulation.

\begin{figure}
	\begin{center}
        \includegraphics[width=.4\columnwidth]{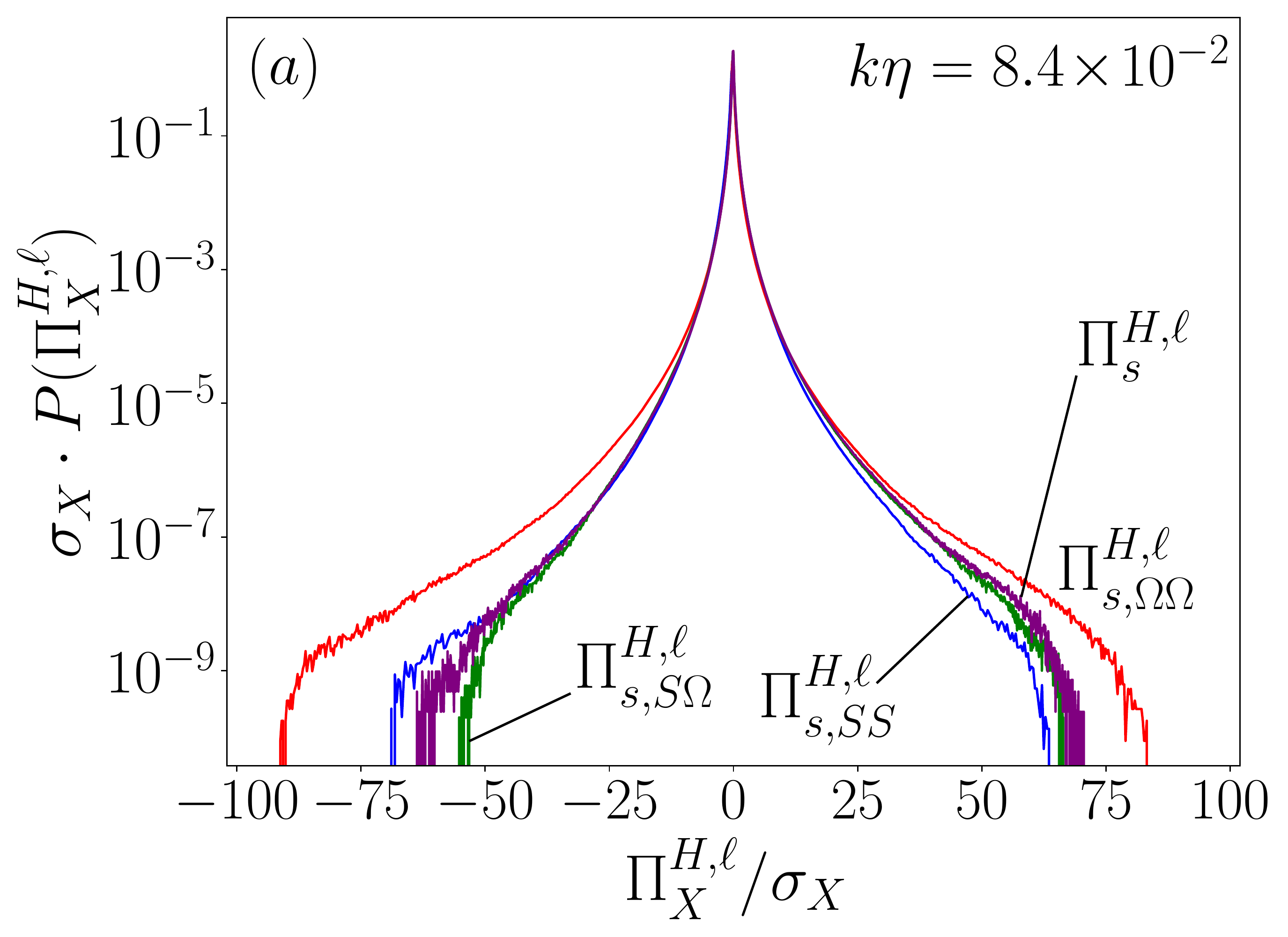}
		\hspace{2em}
        \includegraphics[width=.4\columnwidth]{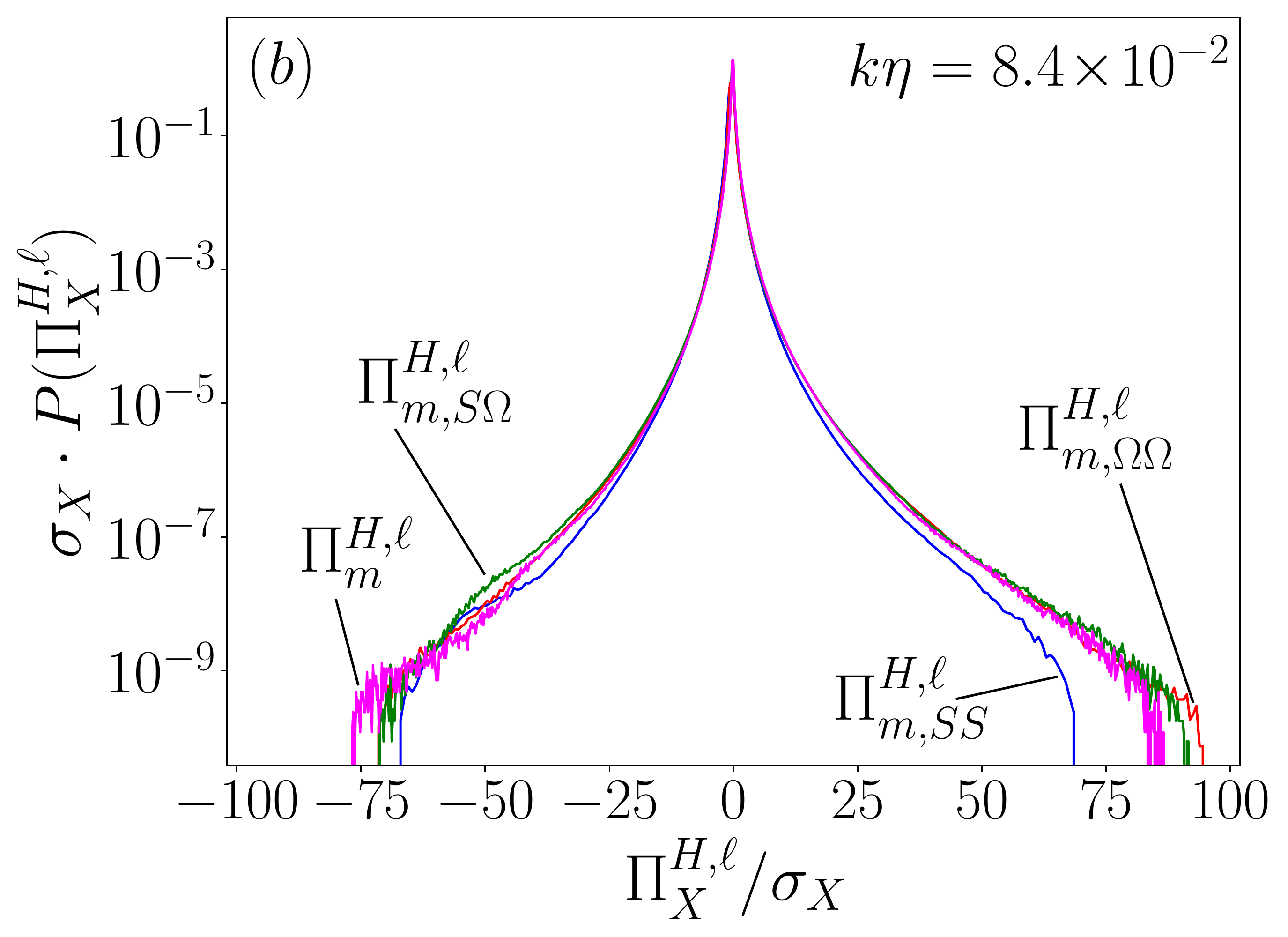}
        \end{center}
        \caption{
		Standardised PDFs of helicity subfluxes $\Pi_X^{H,\ell}$, where $X$ refers to the subflux identifier, for (a) single-scale and (b) multi-scale contributions; $\sigma_X$ denotes the standard deviation of each respective term.
    }
        \label{fig:helicity_pdfs_standard}
\end{figure}

Having discussed the mean subfluxes, we now consider the fluctuations of each subflux term, in order to quantify the level of fluctuations in each term and the presence and magnitude of helicity backscatter. 
Figure \ref{fig:helicity_pdfs_standard} presents
standardised probability density functions (PDFs) of all helicity subfluxes at $k = \pi/\ell = 20$, which is in the inertial range. 
These PDFs are fairly symmetric, much more so than for the kinetic energy fluxes, 
have wide tails, and are strongly non-Gaussian. Single- and multi-scale terms all have strong fluctuations of about 75 standard deviations. 
Interestingly, the subflux term $\Pi_{s,\Omega\Omega}^{H,\ell}$, which necessarily vanishes in mean 
(see~\eqref{eq:SAA_vanish}), has the strongest fluctuations (i.e., is the most intermittent). 
PDFs for all the other subfluxes are comparable. 
The symmetry is more pronounced in the single-scale rather than the multi-scale terms, as can be seen by comparison of the left and right panels of fig.~\ref{fig:helicity_pdfs_standard}.
As all averaged fluxes 
(except $ \langle \Pi_{s, \Omega \Omega}^{H,\ell} \rangle$ which is zero) 
transfer positive helicity from large to small scales, symmetry in the PDFs indicates strong backscatter of positive helicity, or forward scatter of negative helicity. The PDFs become even broader with decreasing filter scale (not shown). 
A comparison between the PDFs of 
$\Pi^{H, \ell}$ and the alternate description based on SGS stresses related to vortex stretching, $\tilde\Pi^{H, \ell}$, has been carried out by \citet{Yan2020}, indicating more intense backscatter in the latter compared to the former. Adding or removing a total gradient can strongly reduce the negative tail of the SGS energy transfer \citep{VelaMartin2022}, and the same may apply to the helicity flux.

\vspace{-1em}
\section{Conclusions} \label{sec:conclusions}
We have derived an exact decomposition of the helicity flux across scales in terms of interactions between vorticity gradients and velocity gradients, and in terms of their scale locality. Decomposing all gradient tensors into symmetric and anti-symmetric parts allows for a discussion and quantification of different physical mechanisms that constitute the helicity cascade. 
Simulation results indicate that all subfluxes transfer helicity from large to small scales, albeit with strong backscatter.
In the inertial range, about 50$\%$ of the total
mean helicity flux 
is due to the action of two scale-local processes: 
(i) vortex flattening 
and
(ii) vortex twisting. 
We have also shown that these two effects are related in mean through a newly derived exact (Betchov-type) relation, which implies that the 
contribution of the former is exactly three times larger than that of the latter.  
Multi-scale effects account for the remaining 50$\%$, with approximate
equipartition between multi-scale versions of the two aforementioned effects and 
multi-scale vortex entangling. 
Thus, it seems likely that, in LES contexts, accurate modeling of the helicity
cascade should not neglect the multi-scale contributions.
Although our numerical quantification of the fluxes is obtained using data from a single simulation with an inertial range of limited length,
we conjecture that the results obtained are robust in the sense that we expect them to hold for flows with larger Reynolds numbers.
Similar flux decompositions can be derived for magnetohydrodynamics. We will report results of these investigations elsewhere in due 
course. 

Computational resources were provided through Scottish Academic Access on Cirrus ({\tt www.cirrus.ac.uk}), 
and the UK Turbulence Consortium 
on ARCHER2 ({\tt www.archer2.ac.uk}).
This work received funding from the European Research Council (ERC) under the
European Union's Horizon 2020 research and innovation programme (grant
agreement No 882340) and from the Priority Programme SPP 1881 ``Turbulent
Superstructures" of the Deutsche Forschungsgemeinschaft (DFG, 
Li3694/1).\\

Competing interests: the authors declare none.
\vspace{-0.5em}
\bibliographystyle{jfm}
\bibliography{references}

\end{document}